\documentclass[aps,twocolumn,groupedaddress,showpacs]{revtex4}

\begin{document}

\title{Quantum retrodiction in open systems}


\author{David T. Pegg}
\email[]{D.Pegg@sct.gu.edu.au}
\affiliation{School of Science, Griffith University, Nathan, Brisbane 4111,
Australia}
\author{Stephen M. Barnett}
\email[]{steve@phys.strath.ac.uk}

\author{John Jeffers}
\email[]{john@phys.strath.ac.uk}
\affiliation{Department of Physics and Applied Physics, University of
Strathclyde, Glasgow G4 0NG, Scotland}

\date{\today}

\begin{abstract}

   Quantum retrodiction involves finding the probabilities for various
   preparation events given a measurement event. This theory
   has been studied for some time but mainly as an interesting
   concept associated with time asymmetry in quantum mechanics.
   Recent interest in quantum communications and cryptography,
   however, has provided retrodiction with a potential practical application.
   For this purpose quantum retrodiction in open systems should be
   more relevant than in closed systems isolated from the environment. In this
   paper we study retrodiction in open systems and develop a general master
   equation for the backward time evolution of the
   measured state, which can be used for calculating preparation
   probabilities. We solve the master equation, by way of example, for the
driven
   two-level atom coupled to the electromagnetic field.
\end{abstract}


\maketitle

\section{Introduction}
The usual formulation of quantum mechanics is predictive. It is well
suited for calculating probabilities for particular outcomes of
measurements from
given preparation information. The preparation information is normally
incorporated into a density operator representing the state of the
prepared system, which evolves until the time of measurement. With
sufficient knowledge we can assign a density operator in the form of
a pure state projector to the system. This contains the maximum amount of
information that nature allows us for prediction. The probability of
the measurement outcome also depends on the operation of the
measurement device.  For a von Neumann measurement \cite{von} a particular
measurement outcome is associated with a pure state projector; for a
more general measurement the outcome is associated with an element of
a probability operator measure (POM) \cite{hel}.  The latter are
non-negative definite operators that sum to the unit operator on the
state space of the system being measured. The retrodictive formalism
of quantum mechanics is used far less often than the predictive
formalism. Originally introduced by Aharonov {\em et al.} \cite{Ahar} in
investigating the origin of the arrow of time, the retrodictive formalism
involves
assigning a state to the system based on knowledge of the measurement
outcome. This state is assigned to the system just prior to the
measurement and evolves backwards in time to the preparation event.
The probability of a particular preparation outcome can be calculated
from a knowledge of the backwards evolved state and the operation of
the preparation device. While the results obtained are fully
consistent with predictive quantum mechanics plus inference based on
Bayes' theorem \cite{usBay}, the retrodictive approach is much more direct
and provides a different insight into quantum mechanics.

Most applications of retrodiction involve closed systems, where the
time evolution is essentially unitary \cite{usBay,lots}. An important
emerging area in which retrodictive quantum mechanics will become
increasingly important is in quantum communication \cite{2},
including quantum cryptography \cite{3}. Here a sender transmits a
system to a recipient after preparing it in a particular state. The
general communication problem is for the recipient to retrodict
the prepared state from the outcome of a measurement and a knowledge
of the operation of the preparation device. For closed systems, the
evolution between preparation and measurement is unitary and the
intrinsic time symmetry significantly simplifies the problem of
calculating the retrodictive evolution. In practice, however,
realistic systems will interact with an environment into which
information is irretrievably lost. This introduces extra uncertainty
in addition to the intrinsic quantum uncertainty associated with the
preparation and measurement and removes the simple time symmetry
associated with unitary evolution in a closed system.  In open
systems with a large environment the predictive evolution can often be
described by a master equation. This describes information loss as the
system propagates forward in time and becomes more entangled with an
environment that is never measured. In this paper we investigate the
derivation of a {\em retrodictive} master equation \cite{PRL} that describes
irreversible loss of information as the system propagates backwards
in time. As will be seen, this is not a simple time inverse of the
predictive master equation.

\section{Preparation and measurement without time delay}

The probabilistic interpretation of quantum mechanics is usually
expressed in terms of the theory of measurement. Retrodiction,
however, relies on preparation being probabilistic so, in this
section, we give a brief overview of the less familiar quantum
theory of preparation and measurement.  To dispense with time evolution
at this stage, we consider an experiment in which one person, the
preparer, prepares a quantum system and then another person, the measurer,
immediately measures it. We assume that there are
readout mechanisms on the preparation and measurement devices that
indicate the preparation event and the measurement event that
occur. The preparer chooses which preparation event will occur and
the measurer chooses whether or not to record the subsequent
measurement event depending on what this measurement event is. If the
measurement event {\em j} is recorded, then so too
is the corresponding preparation event {\em i}. These events are recorded
as a combined event ({\em i, j}). The experiment is repeated
many times with the preparer choosing various states and a list of
recorded combined events is made. The probability associated with a
particular combined event on the list is defined in the usual way as the
suitably normalized occurrence frequency. Both preparer and measurer
have some control over the statistics, which contain elements of
preselection and postselection.

The connection with quantum mechanics can be incorporated in the basic
symmetric expression \cite{prep} for the probability that a combined event
chosen at random on the list is ({\em i, j}):
\begin{equation}
    P^{\Lambda\Gamma}(i,j)=\frac{\text{Tr}(\hat\Lambda_{i}\hat\Gamma_{j})}
    {\text{Tr}(\hat\Lambda\hat\Gamma)}\label{1a}
\end{equation}
where
\begin{equation}
    \hat\Gamma=\sum_{j}\hat\Gamma_{j}\label{2a}
\end{equation}
and
\begin{equation}
    \hat\Lambda=\sum_{i}\hat\Lambda_{i}.\label{3a}
\end{equation}
Here $\hat\Lambda_{i}$ and $\hat\Gamma_{j}$ are non-negative definite
operators associated with the preparation and measurement events {\em i} and
{\em j} respectively. They indicate the state in which the system
is prepared or measured as well as any bias in preparation or in
recording the events
\cite{prep}. The sums in (\ref{2a}) and (\ref{3a}) are over
respectively the measurement and preparation events that can be recorded.
We say that the set of preparation device operators $\hat\Lambda_{i}$
describes mathematically the operation of the preparation device and the set
of measurement device operators $\hat\Gamma_{j}$ describes the operation of
the measurement device. Later, in equation (\ref{A13}), we shall see that
$\hat\Lambda_{i}$ is
$P^{\Lambda}(i)\hat\rho^{\text{pred}}_{i}$ where
$\hat\rho^{\text{pred}}_{i}$ is the density operator for the prepared
state and $P^{\Lambda}(i)$ is the {\em a priori} probability for this
state to be prepared, which reflects any bias in the preparer's
choice. Thus $\hat\Lambda$ is a density operator
representing the best description we can give of the state in which the system
is prepared if we know the operation of the preparation device but
have no knowledge of the preparation or measurement events.
Likewise $\hat\Gamma$ is a density operator
representing the best description we can give of the state in which the system
is measured if we know the operation of the measurement device but
have no knowledge of the preparation or measurement events.

\subsection {Preparation and measurement probabilities}

The basic relation (\ref{1a}) allows us to find expressions for
various probabilities \cite{prep} in
terms of preparation and measurement device operators
$\hat\Lambda_{i}$ and $\hat\Gamma_{j}$. We
use superscripts $\Lambda$ and $\Gamma$ to denote if
the probabilities are based on knowledge of the operations of the
preparation and measurement devices respectively. Where a probability
is based on knowledge of a particular preparation event {\em i} or
measurement event {\em j}, this is shown in the argument of the
probability in the usual way by a vertical stroke preceding {\em i} or
{\em j} respectively. This stroke essentially means `if'. A comma
separating events means `and'. Thus, for example,
$P^{\Lambda\Gamma}(j|i)$ is the probability that measurement event $j$
is recorded if preparation event $i$ is recorded, based on knowledge
of the operations of the preparation and measurement devices.

From the basic relation (\ref{1a}), with (\ref{2a}) and (\ref{3a}), we can
calculate the following probabilities:
\begin{equation}
    P^{\Lambda\Gamma}(i)=\sum_{j}P^{\Lambda\Gamma}(i,j)
    =\frac{\text{Tr}(\hat\Lambda_{i}\hat\Gamma)}
    {\text{Tr}(\hat\Lambda\hat\Gamma)}\label{A1}
\end{equation}
\begin{equation}
    P^{\Lambda\Gamma}(j)=\frac{\text{Tr}(\hat\Lambda\hat\Gamma_{j})}
    {\text{Tr}(\hat\Lambda\hat\Gamma)}\label{A2}
\end{equation}
\begin{equation}
    P^{\Lambda\Gamma}(j|i)=\frac{P^{\Lambda\Gamma}(i,j)}{P^{\Lambda\Gamma}(i)}
    =\frac{\text{Tr}(\hat\Lambda_{i}\hat\Gamma_{j})}
    {\text{Tr}(\hat\Lambda_{i}\hat\Gamma)}\label{A3}
   \end{equation}
\begin{equation}
    P^{\Lambda\Gamma}(i|j)=\frac{\text{Tr}(\hat\Lambda_{i}\hat\Gamma_{j})}
    {\text{Tr}(\hat\Lambda\hat\Gamma_{j})}\label{A4}
\end{equation}

Expression (\ref{A1}) is the probability that, if an experiment
chosen at random has a recorded combined event, this event includes
preparation event {\em i}. Likewise (\ref{A2}) is the probability
that the recorded combined event includes the measurement event
{\em j}.  Expression (\ref{A3}) is the probability that, if the
recorded combined event includes event {\em i}, then it also includes
event {\em j}. Expression (\ref{A3})  can be obtained by limiting the
sample space to those events containing {\em i} and is essentially
Bayes' formula \cite{Boas}.  Likewise (\ref{A4}) is the probability
that the preparation event in a recorded combined event was {\em i}
if the measurement event is {\em j}.

It is clear that we can multiply all the $\hat\Lambda_{i}$ by a
constant value without changing the above probabilities and similarly
for $\hat\Gamma_{j}$. We make use of this flexibility to set
\begin{equation}
    \text{Tr}\hat\Lambda=1\label{A5}
\end{equation}
and
\begin{equation}
    \text{Tr}\hat\Gamma=1\label{A6}
\end{equation}
thus giving these operators the same normalization as density
operators.

Expression (\ref{A3}) can be used for {\em prediction}.  We can calculate
the required probability if we know the preparation device
operator (PDO) $\hat\Lambda_{i}$ associated with the preparation
event {\em i} and if we know the mathematical
description of the operation of the measuring device, that is,
every measurement device operator
(MDO) $\hat\Gamma_{j}$.  Similarly we can
use (\ref{A4}) for {\em retrodiction} from our knowledge of
$\hat\Gamma_{j}$ and all the $\hat\Lambda_{i}$ of the preparation
device.

In the situation where no measurement is made on the system, there is
only one possible measurement outcome and we can speak in terms of an
effective measurement device whose operation is represented by a single
`no-information' MDO that is proportional to the unit operator
\cite{prep}.
Substituting this MDO for $\hat\Gamma_{j}$ in (\ref{A4}) yields
the retrodictive probability $P^{\Lambda\Gamma}(i|j)$, which we write in
this case as $P^{\Lambda}(i)$, where
 \begin{equation}
    P^{\Lambda}(i)=\text{Tr}\hat\Lambda_{i}.\label{A7}
\end{equation}
Expression (\ref{A7}), which depends only on the operation of the
preparation device, is what we would normally call the {\em a priori}
probability for the preparation event to be {\em i}, that is, the
probability in the absence of knowledge of the preparation outcome,
the measurement outcome and the operation of the measurement device.

\subsection{Biased devices}

Consider the special case where the operation of a measurement
device is such that the probability $P^{\Lambda\Gamma}(i)$ that a recorded
preparation event is {\em i} in the absence of knowledge of the
preparation or measurement outcomes is equal to the {\em a priori}
probability $P^{\Lambda}(i)$ for {\em i}.  Then, from (\ref{A1}) and
(\ref{A7}),
\begin{equation}
    \frac{\text{Tr}(\hat\Lambda_{i}\hat\Gamma)}
    {\text{Tr}(\hat\Lambda\hat\Gamma)}=
    \text{Tr}\hat\Lambda_{i}.\label{A8}
\end{equation}
If this is true for all possible preparation devices, that is for all
$\hat\Lambda_{i}$ and $\hat\Lambda$, then from (\ref{A5})
$\hat\Gamma$ must be proportional to the unit operator acting on
the space of the system.  We say that such measurement devices, which
faithfully preserve the {\em a priori} probabilities of preparation
events in the record, have an {\em unbiased} operation. Thus for an
unbiased measuring device operation we can write
\begin{equation}
   \hat\Gamma=G\hat 1\label{A9}
\end{equation}
where $G$ is a positive constant and $\hat 1$  is the unit
operator.  We then define
\begin{equation}
   \hat\Pi_{j}=\hat\Gamma_{j}/G\label{A10}
\end{equation}
which allows us to write (\ref{A3}) as
\begin{equation}
    P^{\Lambda\Gamma}(j|i)
    ={\text{Tr}(\hat\rho^{\text{pred}}_{i}\hat\Pi_{j})}\label{A11}
   \end{equation}
where
\begin{equation}
    \hat\rho^{\text{pred}}_{i}
    =\hat\Lambda_{i}/\text{Tr}\hat\Lambda_{i}.\label{A12}
\end{equation}
This is just the usual, that is predictive, density operator that we assign to
the prepared state on the basis of knowledge of the preparation event.
We note from (\ref{A9}) and (\ref{A10}) that $\hat\Pi_{j}$ are also
non-negative definite and sum to the unit operator.
Thus $\hat\Pi_{j}$ form the elements of a {\em probability operator
measure} (POM) \cite{hel}.  Expression (\ref{A11}), which applies
to unbiased measurement devices, is sometimes regarded as the
fundamental postulate for the probabilistic interpretation of
quantum mechanics \cite{hel}. As we shall see later,
even when the operation of the
measurement device is biased, it is still convenient sometimes to
use $\hat\rho^{\text{pred}}_i$ as defined by (\ref{A12}) and in this case
the predictive probability (\ref{A3}) becomes (\ref{8a}) instead
of (\ref{A11}). Equations (\ref{A12}) and (\ref{A7}) give an
expression for $\hat\Lambda_{i}$ in terms of the predictive
density operator associated with preparation event {\em i} and the
{\em a priori} probability that this state is prepared:
\begin{equation}
    \hat\Lambda_{i}
    =P^{\Lambda}(i)\hat\rho^{\text{pred}}_{i}.\label{A13}
\end{equation}
This allows us to interpret $\hat\Lambda$ in (\ref{3a}) as just the predictive
density operator that we would assign to the prepared state if we knew the
operation of the preparation device but had no knowledge of
the measurement or of the preparation outcome.

Likewise, in the situation where we know only the system state space
and are totally ignorant of the
preparation we are able to represent the operation of the preparation
device by a single `no-information' PDO that is proportional to the
unit operator.  Substituting this PDO for $\hat\Lambda_{i}$
in (\ref {A3}) yields
\begin{equation}
    P^{\Gamma}(j)
    =\text{Tr}\hat\Gamma_{j}.\label{A14}
\end{equation}
We shall call expression (\ref{A14}), which depends only on the
operation of the measurement device, the {\em a priori} probability
for the measurement event to be {\em j}, that is, the probability in
the absence of knowledge of the measurement outcome and any preparation
information.

$P^{\Lambda\Gamma}(j)$ is the probability that a recorded event is
$j$ if the operations of the preparation and measurement devices are
known but the actual preparation event is not known. From (\ref{A2})
and (\ref{A14}) this will be
equal to the {\em a priori} probability for {\em j} if
\begin{equation}
    \frac{\text{Tr}(\hat\Lambda\hat\Gamma_{j})}
    {\text{Tr}(\hat\Lambda\hat\Gamma)}=
    \text{Tr}\hat\Gamma_{j}.\label{A15}
\end{equation}
If this is true for all possible measurement devices, that is for
all $\hat\Gamma_{j}$ and $\hat\Gamma$ , then $\hat\Lambda$
must be
proportional to the unit operator acting on the space of the system.
We say that such preparation devices, which faithfully preserve the
{\em a priori} probabilities of measurement events in the record, have
an {\em unbiased} operation. The retrodictive probability (\ref{A4}) reduces
for an unbiased preparation device operation to
\begin{equation}
    P^{\Lambda\Gamma}(i|j)
    ={\text{Tr}(\hat\rho^{\text{retr}}_{j}\hat\Xi_{i})}.\label{A16}
\end{equation}
where $\hat\Xi_{i}$ sum to the unit operator and are the preparation
POM elements, which are proportional to  $\hat\Lambda_{i}$, and
\begin{equation}
    \hat\rho^{\text{retr}}_{j}
    =\hat\Gamma_{j}/\text{Tr}\hat\Gamma_{j}.\label{A17}
\end{equation}

The symmetry between retrodictive and predictive expressions is evident.
It is worth remarking, however, that the usual experimental situation is
asymmetric.  This arises not from any intrinsic asymmetry in quantum
mechanics but from the asymmetry in the operations of normal preparation
and measurement devices.  The operation of the latter can usually be
described by a POM and so (\ref{A11}) is applicable; the former is
usually not described by a POM so we must use the more general
formula (\ref{A4}) and not (\ref{A16}).   This also results in an
asymmetry between the forms of $P^{\Lambda\Gamma}(i)$ and
$P^{\Lambda\Gamma}(j)$ in that $P^{\Lambda\Gamma}(i)$  is just
the {\em a priori} probability $P^{\Lambda}(i)$  in (\ref{A7}),
which is independent of the operation of the measurement device,
but $P^{\Lambda\Gamma}(j)$  does depend on the operation of the
preparation device.  It is convenient, therefore, to express
$P^{\Lambda\Gamma}(j)$ in terms of $P^{\Lambda}(i)$ .  The expression
for this is
\begin{equation}
    P^{\Lambda\Gamma}(j)
    =\sum_{i}P^{\Lambda\Gamma}(j|i)P^{\Lambda}(i).\label{A18}
\end{equation}
Thus when the operation of the measuring device is
unbiased, Bayes' theorem \cite{Bayes}, for example, can be written in
terms of the {\em a priori} probability $P^{\Lambda}(i)$ as
\begin{equation}
    P^{\Lambda\Gamma}(i|j)=\frac{P^{\Lambda\Gamma}(j|i)P^{\Lambda}(i)}
    {\sum_{i}P^{\Lambda\Gamma}(j|i)P^{\Lambda}(i)}.\label{A19}
\end{equation}

The concept of an unbiased operation of a preparation or measurement
device is of particular relevance for practical cases. For an unbiased
operation of a measurement device, the occurrence frequency of the
preparation event {\em i} in the list of combined events is
proportional to the {\em a priori} probability for {\em i}. An unbiased
measuring device allows the preparer to control the relative occurrence
frequency of a preparation event. Likewise, an unbiased operation of a
preparation device allows the occurrence frequency of the measurement
event {\em j} to be proportional to the {\em a priori} probability for
{\em j}, where this probability is based on knowledge of the operation
of the measurement device but no knowledge of the actual measurement
outcome or of the operation of the preparation device.  An unbiased
preparation device allows the measurer to control the occurrence
frequency of the measurement event. In practice the operations of many,
but not all, measuring devices are unbiased but preparation devices
generally have biased operations. This corresponds to preselection
without postselection. For such cases, as outlined above, we can write
the predictive probability for measurement outcome {\em j} given
preparation event {\em i} in terms of a predictive density operator and
an element of a probability operator measure (POM) for the measuring
device as given by (\ref{A11}). That is, for a measuring device with an
unbiased operation, the preparation device operator $\hat\Lambda_{i}$
converts to a density operator and the measurement device operator
$\hat\Gamma_{j}$ converts to a POM element. We cannot in general,
however, write the corresponding retrodictive probability as the
inverse expression involving an element of a probability operator
measure for the preparation device and a retrodictive density operator.
To do this we would need the operation of the preparation device to be
unbiased, as occurs in some situations such as the Bennett-Brassard
protocol for quantum key distribution \cite {BB84, 3, usBay}.

\section{Time evolution in closed systems}
So far we have considered the preparation time $t_{p}$ to be
the same as the measurement time $t_{m}$ and in (\ref{1a})
we have used $\hat\Lambda_{i}$, the preparation device
operator at $t_{p}$. To allow now for time evolution in a closed
system between $t_{p}$ and $t_{m}$ we replace $\hat\Lambda_{i}$
and $\hat\Lambda$ in the probability (\ref{1a}) by
$\hat\Lambda_{i}(t_{m})$ and $\hat\Lambda(t_{m})$ where
\begin{equation}
    \hat\Lambda_{i}(t_{m})=\hat U(t_{m},t_{p})
    \hat\Lambda_{i}\hat U^{\dagger}(t_{m},t_{p})\label{4a}
\end{equation}
and a corresponding expression for $\hat\Lambda(t_{m})$. Here
$\hat U(t_{m},t_{p})$ is the unitary time-shift
operator. We might interpret this replacement as a modification of 
the original operation of our preparation device to include an
evolutionary period so that the new preparation device operators
are $\hat\Lambda_{i}(t_{m})$.
 The resulting probability expressions obtained from
(\ref{1a}) will
be modified accordingly. Thus the predictive expression (\ref{A3})
becomes
\begin{equation}
    P^{\Lambda\Gamma}(j|i)=\frac{\text{Tr}[\hat U(t_{m},t_{p})
    \hat\Lambda_{i}\hat U^{\dagger}(t_{m},t_{p})
    \hat\Gamma_{j}]}{\text{Tr}[\hat U(t_{m},t_{p})
    \hat\Lambda_{i}
    \hat U^{\dagger}(t_{m},t_{p})\hat\Gamma]}\label{5a}
\end{equation}
and the retrodictive expression (\ref{A4}) becomes
\begin{equation}
    P^{\Lambda\Gamma}(i|j)=\frac{\text{Tr}[\hat U(t_{m},t_{p})
    \hat\Lambda_{i}\hat U^{\dagger}(t_{m},t_{p})
    \hat\Gamma_{j}]}{\text{Tr}[\hat U(t_{m},t_{p})
    \hat\Lambda\hat U^{\dagger}(t_{m},t_{p})
    \hat\Gamma_{j}]}.\label{6a}
\end{equation}

From (\ref{A12}), noting that $\text{Tr}\hat\Lambda_{i}$
is unchanged by a unitary transformation, we see
(\ref{4a}) is equivalent to the usual time development expression
\begin{equation}
    \hat\rho^{\text{pred}}_{i}(t_{m})
    =\hat U(t_{m},t_{p})
    \hat\rho^{\text{pred}}_{i}(t_{p})U^{\dagger}(t_{m},t_{p})\label{7a}
\end{equation}
where $\hat\rho^{\text{pred}}_{i}(t_{p})$
is the usual, that is
predictive, density operator for the prepared state at $t_{p}$. Using
(\ref{A12}) we can rewrite (\ref{5a}) as
\begin{equation}
    P^{\Lambda\Gamma}(j|i)=\frac{\text{Tr}[\hat\rho^{\text{pred}}_{i}(t_{m})
    \hat\Gamma_{j}]}{\text{Tr}[\hat\rho^{\text{pred}}_{i}(t_{m})
    \hat\Gamma]}.\label{8a}
\end{equation}
A natural interpretation of (\ref{8a}) is in terms of the system being
prepared in the state $\hat\rho^{\text{pred}}_{i}(t_{p})$ at $t_{p}$
and then evolving forwards in time to become
$\hat\rho^{\text{pred}}_{i}(t_{m})$ at the
measurement time.

At first sight, it may appear that the evolution has destroyed the
symmetry between (\ref{A3}) and (\ref{A4}) which have now become
(\ref{5a}) and (\ref{6a}). However by using the cyclic property of the
trace we rewrite both (\ref{5a}) and (\ref{6a}) in terms of
$\hat\Gamma_{j}(t_{p})$ where
\begin{equation}
    \hat\Gamma_{j}(t)=\hat U^{\dagger}(t_{m},t)
    \hat\Gamma_{j}\hat U(t_{m},t),\label{9a}
\end{equation}
with $t_{p}\leq t\leq t_{m}$. Then, for example, the retrodictive
formula (\ref{6a}) becomes
\begin{equation}
    P^{\Lambda\Gamma}(i|j)=\frac{\text{Tr}[\hat\Lambda_{i}
    \hat\Gamma_{j}(t_{p})]}{\text{Tr}[
    \hat\Lambda\hat \Gamma_{j}(t_{p})]}.\label{10a}
\end{equation}
Expressions (\ref{9a}) and (\ref{10a}) can be interpreted in terms of
a new operation of the measurement device which incorporates a prior
time delay as part of the measurement process.  This operation
is described by the set of measurement device operators
$\hat\Gamma_{j}(t_{p})$.

We can use (\ref{9a}) and (\ref{A17}) to write (\ref{10a}) as
\begin{equation}
    P^{\Lambda\Gamma}(i|j)=\frac{\text{Tr}[\hat\Lambda_{i}
    \hat\rho^{\text{retr}}_{j}(t_{p})]}{\text{Tr}[
    \hat\Lambda\hat\rho^{\text{retr}}_{j}(t_{p})]}\label{11a}
\end{equation}
where
\begin{equation}
    \hat\rho^{\text{retr}}_{j}(t_{p})=\hat U^{\dagger}(t_{m},t_{p})
    \hat\rho^{\text{retr}}_{j}(t_{m})\hat
    U(t_{m},t_{p}).\label{12a}
\end{equation}
Expressions (\ref{11a}) and (\ref{12a}) can be interpreted as the
system being assigned a retrodictive state
$\hat\rho^{\text{retr}}_{j}(t_{m})$ at time $t_{m}$ on the basis of
the recorded measurement outcome and evolving backwards in time to the
preparation time $t_{p}$.

Any of the above four interpretations of the process of preparation
and measurement can be applied to both the
predictive and retrodictive probabilities (\ref{5a}) and (\ref{6a}).
It is not necessary, for example, to use the predictive density
operator for prediction, nor the retrodictive density operator for
retrodiction. Use of the forward-evolved predictive density
operator is convenient, however, when we know the outcome of a
particular preparation and wish to retain flexibility in calculating
probabilities for a range of possible measurement events with
different measuring devices. Likewise, if we know the measurement
outcome and wish to have flexibility in calculating probabilities for
a range of possible preparation events with different preparation
devices, it is convenient to calculate the backward-evolved
retrodictive density operator. This is the purpose of this paper,
except instead of finding the unitary evolution for a closed system,
we wish to study the non-unitary evolution of a more realistic open
system.

\section{Open systems}
In a preliminary paper \cite{PRL} we outlined the derivation of
an open system retrodictive master equation for the particular case
where the operation of the measurement device can be described
by a POM. In this section we examine in detail the more general
case where the set of (renormalized) measurement
device operators may not necessarily form a POM and also give a proof of the
non-negativity of the retrodictive density operator at all times
between measurement and preparation. Later we solve the master
equation for a driven two-level atom.

We are interested in a quantum system {\em S} interacting with an
environment {\em E}. The initial state of the environment at time
$t_{p}$ is known but
the state of the environment subsequent to this is never measured.
Before examining the retrodictive problem we outline first the
usual predictive problem. In the predictive problem, in addition to the
initial state of {\em E}, we also know the initial state of {\em S}
and wish to predict the probabilities for the outcomes {\em j} of possible
later measurements of {\em S}. For this we can use expressions (\ref{8a})
and (\ref{7a}) where $\hat\rho^{\text{pred}}_{i}(t_{p})$ is now the
predictive density operator for the combined system of {\em E} and {\em
S}, which will be the product of the initial density operators for
{\em E} and {\em S}:
\begin{equation}
    \hat\rho^{\text{pred}}_{i}(t_{p})
    =\hat\rho^{\text{pred}}_{E}\otimes
    \hat\rho^{\text{pred}}_{i,S}.
    \label{13a}
\end{equation}
This initial state will evolve to a state which is in general entangled
in accord with (\ref{7a}) where the unitary operator will now
involve the Hamiltonian
of the combined system. Each measurement device operator will be a
product an operator $\hat\Gamma_{j,S}$ acting on {\em S}
and an operator acting on the environment. As discussed earlier,
the measurement device operator for a non-measurement is
proportional to the unit operator as this is effectively an
unbiased measurement with only one possible outcome \cite{prep}. Thus the
predictive probability (\ref{8a}) becomes for this case
\begin{equation}
P^{\Lambda\Gamma}(j|i)=\frac{\text{Tr}_{ES}[\hat\rho^{\text{pred}}_{i}(t_{m})
\hat\Gamma_{j,S}\otimes\hat1_{E}]}{
    \text{Tr}_{ES}[\hat\rho^{\text{pred}}_{i}(t_{m})
    \hat\Gamma_{S}\otimes\hat1_{E}]}.\label{14a}
\end{equation}
where $\hat 1_{E}$ is the unit operator on the space of the the
environment and $\hat\Gamma_{S}$ is the sum of
$\hat\Gamma_{j,S}$. The trace is over the environment and the
system {\em S} states.
By defining a reduced predictive density operator for the system {\em S} at
the time {\em t} as \cite {book}
\begin{equation}
    \hat\rho^{\text{pred}}_{i,S}(t)=\text{Tr}_{E}[\hat
    U(t,t_{p})\hat\rho^{\text{pred}}_{E}\otimes
    \hat\rho^{\text{pred}}_{i,S}\hat
    U^{\dagger}(t,t_{p})]\label{15a}
\end{equation}
we can write the predictive probability in terms of operators acting
on the state space of {\em S}. Substituting (\ref{7a}), (\ref{13a})
and (\ref{15a}) with $t = t_{m}$ into (\ref{14a}) gives
\begin{equation}
    P^{\Lambda\Gamma}(j|i)=\frac{\text{Tr}_{S}
    [\hat\rho^{\text{pred}}_{i,S}(t_{m})
    \hat\Gamma_{j,S}]}{
    \text{Tr}_{S}[\hat\rho^{\text{pred}}_{i,S}(t_{m})
    \hat\Gamma_{S}]},\label{16a}
\end{equation}
which is of the same form as the closed system formula (\ref{8a}).
Thus in order to find the predictive probability
$P^{\Lambda\Gamma}(j|i)$ we need only calculate the evolution of the
reduced predictive density operator rather than the evolution of the
combined system plus environment. If the environment
{\em E} has a large number of degrees of freedom and is little
changed by the coupling to {\em S}, then if the environment is not
measured we can make use of the
approximation that, for any time {\em t} between preparation and
measurement \cite {book},
\begin{equation}
    \hat
    U(t,t_{p})\hat\rho^{\text{pred}}_{E}\otimes
    \hat\rho^{\text{pred}}_{i,S}\hat
    U^{\dagger}(t,t_{p})
    \approx \hat\rho^{\text{pred}}_{E}\otimes
    \hat\rho^{\text{pred}}_{i,S}(t).\label{17a}
\end{equation}
The Markov approximation \cite {book} then leads to a general Markovian master
equation for $\hat\rho^{\text{pred}}_{i,S}(t)$ that, in standard
Lindblad form, is given by \cite{11}
\begin{eqnarray}
    \dot{\hat\rho}^{\text{pred}}_{i,S}(t)&=&-i{\hbar}^{-1}
    [\hat H_{S},\hat\rho^{\text{pred}}_{i,S}(t)] \nonumber
    \\&&\mbox{}+\sum_{q}[2\hat A_{q}\hat\rho^{\text{pred}}_{i,S}(t)
    \hat A^{\dagger}_{q}-\hat A^{\dagger}_{q}
    \hat A_{q}\hat\rho^{\text{pred}}_{i,S}(t)\nonumber
    \\&&\mbox{}
    -\hat\rho^{\text{pred}}_{i,S}(t)\hat A^{\dagger}_{q}\hat
    A_{q}]\label{18a}
\end{eqnarray}
where $\hat H_{S}$ is the Hamiltonian for the system {\em S} without
the environment and $\hat A_{q}$ is a system operator.
This general form of equation incorporates the Markovian requirement
and conserves both the non-negative definiteness
and the trace of the reduced density operator \cite {11}.

The expression for the retrodictive probability (\ref {10a}) is
\begin{equation}
    P^{\Lambda\Gamma}(i|j)=\frac{\text{Tr}_{ES}[
    \hat\Lambda_{i,S}\otimes\hat\rho^{\text{pred}}_{E}
    \hat\Gamma_{j}(t_{p})]}
    {\text{Tr}_{ES}[\hat\Lambda_{S}
    \otimes\hat\rho^{\text{pred}}_{E}
    \hat\Gamma_{j}(t_{p})]}
    \label{19a}
\end{equation}
where the backward-evolved measurement device operator
$\hat\Gamma_{j}(t)$ is given by (\ref {9a}),
$\hat\Lambda_{S}$ is the sum of $\hat\Lambda_{i,S}$ and
we have set the proportionality constant between the
environmental preparation device operator and
$\hat\rho^{\text{pred}}_{E}$
to unity because this is the {\em a priori} probability that the environment
is prepared in this state. For our case, where the environment is
not measured, the measurement device operator
$\hat\Gamma_{j}$ will be proportional to
$\hat\Gamma_{j,S}\otimes\hat1_{E}$. As $\hat\Gamma_{j}(t)$
evolves backwards in time towards $t_{p}$, it changes from being
factorizable at $t_{m}$ to become in general more entangled.

We seek to express $P^{\Lambda\Gamma}(i|j)$ in terms of a {\em reduced
retrodictive density operator} so that it has a form similar to
(\ref{11a}), just as $P^{\Lambda\Gamma}(j|i)$ in (\ref{16a}) is
similar to (\ref{8a}). The simplest way to do this is first to define
a reduced backward evolved measurement device operator and
then to normalize this so that the trace is unity. From (\ref{19a})
we see that we need a reduced backward-evolved measurement device
operator in the form
\begin{equation}
    \hat\Gamma_{j,S}(t) \propto {\text{Tr}_{E}}
    [\hat\rho^{\text{pred}}_{E}\hat U^{\dagger}(t_{m},t)
    \hat\Gamma_{j,S}\otimes\hat1_{E}
    \hat U(t_{m},t)].\label{20a}
\end{equation}
We can now
define a reduced retrodictive density operator for the system
{\em S} at the time {\em t} as
\begin{equation}
    \hat\rho^{\text{retr}}_{j,S}(t)=
    \frac{\hat\Gamma_{j,S}(t)}
    {\text{Tr}_{S}[\hat\Gamma_{j,S}(t)]}
    \label{21a}
\end{equation}
Then by substituting (\ref{21a}) and (\ref{20a}) into
(\ref{19a}) we can write the retrodictive probability in terms of
operators acting on the state space of {\em S}:
\begin{equation}
    P^{\Lambda\Gamma}(i|j)=\frac{\text{Tr}_{S}[
    \hat\Lambda_{i,S}\hat\rho^{\text{retr}}_{j,S}(t_{p})]}
    {\text{Tr}_{S}[\hat\Lambda_{S}
    \hat\rho^{\text{retr}}_{j,S}(t_{p})]}.
    \label{22a}
\end{equation}
Thus, if we know the reduced retrodictive density operator and the
operation of the preparation device, we can calculate the retrodictive
preparation probabilities by a formula of the same form (\ref{11a})
as for a closed system. For this, we wish to find an appropriate
master equation governing the evolution backwards in time. We
should note that the reduced predictive and retrodictive density
operators in (\ref{15a}) and (\ref{21a}) are not simply related to
each other and thus the master equation for the reduced retrodictive
density operator is not immediately derivable from (\ref{18a}),
for example simply by reversing the sign of the time. This, of
course, is not due to some intrinsic time asymmetry in quantum
mechanics; it arises from the asymmetry in the
environmental boundary conditions. For both the predictive and
retrodictive cases we are assuming that the environment is prepared in
some particular state but not measured. To obtain symmetry we would
have to assume in the retrodictive case that the environment is
measured to be in some state and that we have no information about its
preparation.

The most straightforward way of finding a retrodictive master equation
for $\hat\rho^{\text{retr}}_{j,S}(t)$ is to find an equation for the
backwards evolution of $\hat\Gamma_{j,S}(t)$ and then use
(\ref{21a}) to obtain the master equation. To do this,  we use
(\ref{7a}), the group property of the time-shift operator
\begin{equation}
   \hat U(t_{m},t_{p})=\hat U(t_{m},t)
   \hat U(t,t_{p}),\label{23a}
\end{equation}
where $t$ is a time between $t_{m}$ and $t_{p}$, and the cyclic
property of the trace to rewrite the numerator of the right side of
(\ref{14a}) as
\begin{eqnarray}
   \text{Tr}_{ES}[ \hat U(t,t_{p})
   \hat\rho^{\text{pred}}_{E}\otimes
    \hat\rho^{\text{pred}}_{i,S}
   \hat U^{\dagger}(t,t_{p})\hat
   U^{\dagger}(t_{m},t)\hat\Gamma_{j,S}\nonumber
 \\
   \otimes\hat1_{E}\hat U(t_{m},t)].\label{24a}
\end{eqnarray}
We leave the denominator of the right side of (\ref{14a}) as it is.
Clearly $P^{\Lambda\Gamma}(j|i)$ does not vary as we change $t$
between $t_{p}$ and $t_{m}$ and so the derivative of the numerator
(\ref{24a}) with respect to $t$ must vanish. Using the weak-coupling
approximation (\ref{17a}) and (\ref{20a}) we can write (\ref{24a}) as
proportional to
\begin{equation}
   \text{Tr}_{S}[\hat\rho^{\text{pred}}_{i,S}(t)
   \hat\Gamma_{j,S}(t)].
   \label{25a}
\end{equation}
The vanishing of the derivative of this with respect to $t$ yields
\begin{equation}
    \text{Tr}_{S}[\hat\rho^{\text{pred}}_{i,S}(t)
   \dot{\hat\Gamma}_{j,S}(t)]=
   -\text{Tr}_{S}[\dot{\hat\rho}^{\text{pred}}_{i,S}(t)
   \hat\Gamma_{j,S}(t)].\label{26a}
\end{equation}
Substituting (\ref{18a}) into (\ref{26a}) gives, after application of
the cyclic property of the trace,
\begin{eqnarray}
    \text{Tr}_{S}[\hat\rho^{\text{pred}}_{i,S}(t)
   \dot{\hat\Gamma}_{j,S}(t)]&=&
   \text{Tr}_{S}(\hat\rho^{\text{pred}}_{i,S}(t)
   \{-i\hbar^{-1}[\hat H_{S},\hat\Gamma_{j,S}(t)]\nonumber
\\
   &&\mbox{}-\sum_{q}[2\hat A^{\dagger}_{q}\hat\Gamma_{j,S}(t)
   \hat A_{q}\nonumber
\\&&\mbox{}-
   \hat\Gamma_{j,S}(t)A^{\dagger}_{q}\hat A_{q}\nonumber
   \\&&\mbox{}
    -\hat A^{\dagger}_{q}\hat A_{q}\hat\Gamma_{j,S}(t)]\}).
    \label{27a}
\end{eqnarray}
This is true for all $\hat\rho^{\text{pred}}_{i,S}(t)$, so the
evolution equation for the reduced measurement device operator is
\begin{eqnarray}
    \dot{\hat\Gamma}_{j,S}(t)&=&
   -i\hbar^{-1}[\hat H_{S},\hat\Gamma_{j,S}(t)]\nonumber
\\
   &&-\sum_{q}[2\hat A^{\dagger}_{q}\hat\Gamma_{j,S}(t)
   \hat A_{q} \nonumber
\\&&\mbox{}
   -\hat\Gamma_{j,S}(t)A^{\dagger}_{q}\hat A_{q}
   -\hat A^{\dagger}_{q}\hat A_{q}\hat\Gamma_{j,S}(t)].\label{28a}
\end{eqnarray}
We can then find the retrodictive master equation for
$\hat\rho^{\text{retr}}_{j,S}(t)$ by substituting (\ref{28a}) into the
time derivative of (\ref{21a}). This gives
\begin{eqnarray}
    \dot{\hat\rho}^{\text{retr}}_{j,S}(t)&=&
   -i\hbar^{-1}[\hat H_{S},\hat\rho^{\text{retr}}_{j,S}(t)]\nonumber
\\
   &&-\sum_{q}[2\hat A^{\dagger}_{q}\hat\rho^{\text{retr}}_{j,S}(t)
   \hat A_{q}-
   \hat\rho^{\text{retr}}_{j,S}(t)A^{\dagger}_{q}\hat A_{q} \nonumber
\\
   &&\mbox{}-\hat A^{\dagger}_{q}\hat
   A_{q}\hat\rho^{\text{retr}}_{j,S}(t)]\nonumber
\\
 &&\mbox{}-2\hat\rho^{\text{retr}}_{j,S}(t)\text{Tr}_{S}
 \{\hat\rho^{\text{retr}}_{j,S}(t)\sum_{q}[\hat A^{\dagger}_{q},
 \hat A_{q}]\}\label{29a}
\end{eqnarray}
as the master equation for $\hat\rho^{\text{retr}}_{j,S}(t)$.

It is not difficult to see that (\ref{29a}) conserves the trace of
$\hat\rho^{\text{retr}}_{j,S}(t)$, but (\ref{28a}) does not preserve
the trace of $\hat\Gamma_{j,S}(t)$. The price of this preservation
is that while (\ref{28a}) is a linear differential
equation, the master equation (\ref{29a}) is more complicated.
While (\ref{29a}) can be solved directly for some simple cases,
the solution is not always obvious. In general, however, (\ref{29a})
is reducible to a linear equation by substituting a
variable-trace operator
\begin{equation}
    \hat B(t) =\hat\rho^{\text{retr}}(t)\exp \left( 2\int_{t_{m}}^{t}
    \text{Tr}\{\hat\rho^{\text{retr}}(t')
    \sum_{q}[\hat A^{\dagger}_{q},\hat A_{q}]\}dt'\right),
    \label{30a}
\end{equation}
which gives a linear equation for $\hat B(t)$. Then
$\hat\rho^{\text{retr}}(t)$ can be recovered from the solution as
$\hat B(t)/\text{Tr}\hat B(t)$. The linear equation obtained,
however, is the same as (\ref{28a}) so $\hat B(t)$ is just
proportional to $\hat\Gamma_{j,S}(t)$. It is thus more convenient
in general to calculate retrodictive probabilities by solving (\ref{28a})
for $\hat\Gamma_{j,S}(t)$ and then using
\begin{equation}
    P^{\Lambda\Gamma}(i|j)=\frac{\text{Tr}_{S}[
    \hat\Lambda_{i,S}\hat\Gamma_{j,S}(t_{p})]}
    {\text{Tr}_{S}[\hat\Lambda_{S}
    \hat\Gamma_{j,S}(t_{p})]},
    \label{31a}
\end{equation}
which is derivable from (\ref{20a}),(\ref{19a}) and (\ref{9a}),
instead of solving (\ref{29a}) for $\hat\rho^{\text{retr}}_{j,S}(t)$
and using (\ref{22a}).

To give a physical interpretation of
$\hat\Gamma_{j,S}(t)$, we first show that this is a
non-negative operator for all $t$ between $t_{p}$ and $t_{m}$ as
follows. Consider an operator $\hat\sigma(t)$, acting on the state
space of {\em S}, defined to obey the predictive master equation
(\ref{18a}) for times $t$ between (and including) $T$ and $t_{m}$
where $T$ is some time such that $t_{p}\leq T\leq t_{m}$. Let
$\hat\sigma(T)$ = $|u\rangle\langle u|$ where
$|u\rangle$ is some arbitrary pure state of $S$. Then $\hat\sigma(t)$
will have the form of a predictive density operator and so
$\hat\sigma(t_{m})$ will be non-negative definite. From (\ref{20a}),
$\hat\Gamma_{j,S}(t_{m})$ is just $\hat\Gamma_{j,S}$, which
is also non-negative definite. Thus
\begin{equation}
    \text{Tr}_{S}[\hat\sigma(t_{m})\hat\Gamma_{j,S}(t_{m})]\geq 0.
    \label{32a}
\end{equation}
It is not difficult to show directly  from (\ref{18a}) with
$\hat\sigma(t)$ in place of $\hat\rho^{\text{pred}}_{i,S}(t)$ and from
(\ref{28a}) that the time derivative of
$\text{Tr}_{S}[\hat\sigma(t)\hat\Gamma_{j,S}(t)]$
is zero for $t$
between $T$ and $t_{m}$. Consequently
\begin{equation}
    \text{Tr}_{S}[\hat\sigma(T)\hat\Gamma_{j,S}(T)]\geq 0.
    \label{33a}
\end{equation}
Writing the left side of (\ref{33a}) as
$\langle u|\hat\Gamma_{j,S}(T)|u\rangle$ shows that
$\hat\Gamma_{j,S}(T)$ is non-negative definite for all times $T$
between $t_{p}$ and $t_{m}$.

The non-negativity of $\hat\Gamma_{j,S}(t)$ has two consequences.
Firstly it follows from (\ref{21a}) that
$\hat\rho^{\text{retr}}_{j,S}(t)$ is also non-negative. This,
combined with the trace of $\hat\rho^{\text{retr}}_{j,S}(t)$ being
conserved as unity, means that (\ref{29a}) is a legitimate master
equation. The second consequence is that $\hat\Gamma_{j,S}(t)$
can be considered to be measurement device operators associated with
measurement events $j$. Furthermore,
if $\hat\Gamma_{j,S}(t)$ sum to be proportional to the unit
operator on the state space of $S$ it follows from (\ref{28a}) that
this sum is also conserved. Thus if the set of operators
$\hat\Gamma_{j,S}(t_{m})$, which are proportional to
$\hat\Gamma_{j,S}\otimes \hat 1_{E}$, describes the operation of an
unbiased measuring device, so too will the set of operators
$\hat\Gamma_{j,S}(t)$ and thus also $\hat\Gamma_{j,S}(t_{m})$.
Consequently the open system considered here allows
a similar interpretation as does the closed system equation
(\ref{10a}) with the measurement, or collapse of the state, taking
place immediately after the preparation time $t_{p}$.
It should be noted that in the common interpretation, in which the
operation of the measurement device is
described by $\hat\Gamma_{j,S}(t_{m})$, the initial environment
state is considered to be part of the description of the operation of
the preparation device. For the new interpretation, with the
measurement taking place at $t_{p}$ in accord with (\ref {31a}),
the initial environment
state is considered to be part of the description of the operation of
the measurement device. This is analogous to a homodyne detection
system in which the initial state of the local oscillator is considered
to be part of the operation of the measurement device rather than
part of the operation of the preparation device along with the signal
field state. The conservation of non-negativity allows us, in fact,
to interpret (\ref{8a}) and (\ref{10a}) in terms of the measurement or
collapse taking place at {\em any} time $t$ between $t_{p}$ and $t_{m}$ with
non-negative preparation and measurement device operators defined
appropriately. The physical interpretation of the invariance of
$P^{\Lambda\Gamma}(j|i)$ under changes of $t$ is that measurable
probabilities are independent of when we choose the collapse time.
This underlines the somewhat arbitrary nature of this concept.

By comparing the predictive and retrodictive master equations
(\ref{18a}) and (\ref{29a}), we see that when the interaction with
the environment can be ignored, for example for short enough time
intervals, the evolution is given by just the first term in each. The
two equations for this unitary evolution are then the same. It should
be remembered, however, that we have defined the time $t$ so that it
increases in both cases. For retrodiction, it is more natural to
define a premeasurement time as
\begin{equation}
    \tau=t_{m}-t. \label{34a}
\end {equation}
so that
\begin{equation}
    d\hat\rho^{\text{retr}}_{_{j,S}}(\tau)/d\tau=
    -\dot{\hat\rho}^{\text{retr}}_{_{j,S}}(t). \label{35a}
\end {equation}
The unitary part of the retrodictive master equation in terms of
$\tau$ is then seen explicitly to be the time inverse of the
corresponding part of the predictive master equation. That the
complete retrodictive master equation is not simply the time inverse
of the predictive equation is illustrated by the following. If
there is an extremely long time between preparation and measurement,
both $\hat\rho^{\text{pred}}_{_{j,S}}(t)$ and
$\hat\rho^{\text{retr}}_{_{j,S}}(\tau)$ can reach their steady-state
values for which $\dot{\hat\rho}^{\text{pred}}_{_{j,S}}(t)$
and $d\hat\rho^{\text{retr}}_{_{j,S}}(\tau)/d\tau$ are zero. A
solution of the resulting steady-state retrodictive equation is
clearly $\hat\rho^{\text{retr}}_{_{j,S}}(\infty)$ =
$\hat 1_{S}/D$ where $\hat 1_{S}$ is the unit operator acting on the
state space of $S$ and $D$ is the dimension of that space. This
represents the no-information state. Essentially, at the measurement
time the retrodictive state is the product of the measured state and
the no-information environment state. As we go backwards
in time from the measurement, the retrodictive system state becomes
more entangled with the environment and we lose information about the
system state. In the limit of very long times in the past, the system
state becomes completely unretrodictable. There is no similar simple
general solution of the predictive master equation, however, which
would imply unpredictability for long times. Indeed for the case of
an excited two-level atom undergoing spontaneous emission into the
environmental vacuum field, for example, it becomes
very likely that the atom will be found in its ground state in the long
term future, an outcome which is very predictable.

\section{Coherently driven atom}
\subsection{Retrodictive density operator}

As an example of a retrodictive master equation and its solution, we
examine the case of a two-level atom driven on resonance by a strong
laser field. The solution of the predictive master equation for this
system is well known \cite{book}, with the atom exhibiting damped Rabi
oscillations at a frequency $\Omega$ = $(V^{2}-\gamma^{2}/4)^{1/2}$
where $V$ is proportional to the strength of the laser field and
$\gamma$ is the spontaneous decay rate. We can write the Hamiltonian
describing the interaction between the atom and the laser field in
the interaction picture in the semiclassical form
\begin{equation}
    \hat H=
    \frac{\hbar}{2}V(\hat\sigma_{+}+\hat\sigma_{-}) \label{36a}
\end {equation}
where $\sigma_{+}$ = $|e\rangle\langle g|$ and $\sigma_{-}$ =
$|g\rangle\langle e|$. A comparison of (\ref{18a}) with the predictive
master equation for this system \cite{book}, that is,
\begin{eqnarray}
    \dot{\hat\rho}^{\text{pred}}_{i,S}(t)&=&
    -\frac{iV}{2}[(\hat \sigma_{+}+\hat
    \sigma_{-}),\hat\rho^{\text{pred}}_{i,S}(t)]\nonumber
    \\
    &&+\gamma[2\hat \sigma_{-}\hat\rho^{\text{pred}}_{i,S}(t)
    \hat \sigma_{+}-\hat \sigma_{+}
    \hat \sigma_{-}\hat\rho^{\text{pred}}_{i,S}(t) \nonumber
 \\
    &&\mbox{}-\hat\rho^{\text{pred}}_{i,S}(t)\hat \sigma_{+}\hat
\sigma_{-}],\label{36b}
\end{eqnarray}
shows that $q$ in (\ref{18a})
has only one value, with $A_{q}$ = $\gamma^{1/2}\sigma_{-}$.
Thus, using
$\hat B(\tau)$ in place of $\hat\Gamma_{j,S}(\tau)$, we can write the linear
form (\ref{28a}) of the retrodictive master equation
in terms of the premeasurement time $\tau$ as
\begin{eqnarray}
    d\hat B(\tau)/d\tau&=&
   \frac{iV}{2}[(\hat\sigma_{+}+\hat\sigma_{-}),\hat B(\tau)]\nonumber
\\
   &&+\gamma[2\hat\sigma_{+}\hat B(\tau)
   \hat\sigma_{-}-
   \hat B(\tau)\hat\sigma_{+}\hat\sigma_{-}\nonumber
\\
   &&\mbox{}-\hat\sigma_{+}\hat\sigma_{-}\hat B(\tau)].\label{37a}
\end{eqnarray}

We can solve (\ref{37a}) by converting it to a set of c-number equations.
We write
\begin{equation}
    \hat
B(\tau)=u(\tau)\hat\sigma_{1}+v(\tau)\hat\sigma_{2}+w(\tau)\hat\sigma_{3}
    +x(\tau)\hat1_{S}\label{38a}
\end {equation}
where $\hat1_{S}$ is the unit operator acting on the space of the atom and
\begin{equation}
    \hat\sigma_{1}=\hat\sigma_{+}+\hat\sigma_{-}\label{39a}
\end{equation}
\begin{equation}
   \hat\sigma_{2}=-i(\hat\sigma_{+}-\hat\sigma_{-})\label{40a}
\end{equation}
\begin{equation}
   \hat\sigma_{3}=2\hat\sigma_{+}\hat\sigma_{-}-\hat 1_{S}.\label{41a}
\end{equation}
Then substitution of (\ref{38a}) into (\ref{37a}) yields
\begin{eqnarray}
   \lefteqn{\frac{du}{d\tau}\hat\sigma_{1}+\frac{dv}{d\tau}\hat\sigma_{2}+
   \frac{dw}{d\tau}\hat\sigma_{3}+\frac{dx}{d\tau}\hat 1_{S}=}\nonumber
   \\
   &&-\gamma u\hat\sigma_{1}-(\gamma v-Vw)\hat\sigma_{2}
   -(Vv+{2\gamma w)}\hat\sigma_{3}-2\gamma w\hat 1_{S}.\label{42a}
\end {eqnarray}
Multiplying (\ref{42a}) by $\hat\sigma_{1}$,
$\hat\sigma_{2}$ and $\hat\sigma_{3}$ respectively and taking the
trace of both sides gives
\begin{equation}
   \frac{du}{d\tau}=-\gamma u\label{43a}
\end{equation}
\begin{equation}
   \frac{dv}{d\tau}=-\gamma v+Vw\label{44a}
\end{equation}
\begin{equation}
   \frac{dw}{d\tau}=-Vv-2\gamma w\label{45a}
\end{equation}
and taking the trace of both sides of (\ref{42a}) gives
\begin{equation}
   \frac{dx}{d\tau}=-2\gamma w.\label{46a}
\end{equation}

The solution of (\ref{43a}) is simply
\begin{equation}
   u(\tau)=u(0)\exp (-\gamma\tau).\label{47a}
\end{equation}
The simultaneous equations (\ref{44a}) and (\ref{45a}) are
straightforwardly solved by standard means \cite{Menzel} to yield
\begin{eqnarray}
   v(\tau)&=&\exp(-3\gamma\tau/2)\{v(0)[\cos(\Omega\tau) \nonumber
   \\&&\mbox{}+\gamma(2\Omega)^{-1}\sin(\Omega\tau)]\nonumber
   \\&&\mbox{}+ w(0)V\Omega^{-1}\sin(\Omega\tau)\}\label{48a}
\end{eqnarray}
and
\begin{eqnarray}
   w(\tau)&=&\exp(-3\gamma\tau/2)\{-v(0)V\Omega^{-1}\sin(\Omega\tau)
   \nonumber
   \\&&\mbox{}+w(0)[\cos(\Omega\tau)
   -\gamma(2\Omega)^{-1}\sin(\Omega\tau)]\}\label{49a}
\end{eqnarray}
which allows us to find $x$ from (\ref{46a}) as
\begin{eqnarray}
   x(\tau)&=&x(0)+\frac{2\gamma[Vv(0)-\gamma w(0)]}{2\gamma^{2}+V^{2}}\nonumber
 \\
    &&
    +\exp(-3\gamma\tau/2)\frac{-2\gamma}{2\gamma^{2}+V^{2}}\nonumber
 \\&&\times
    \{\frac{(\gamma^{2}+2V^{2})w(0)+3\gamma Vv(0)}{2\Omega}
    \sin(\Omega\tau)\nonumber
  \\
    &&+[Vv(0)-\gamma w(0)]\cos(\Omega\tau)\}.\label{50a}
\end{eqnarray}

Substituting (\ref{47a}), (\ref{48a}), (\ref{49a}) and (\ref{50a})
into (\ref{38a}) gives $\hat B(\tau)$. The retrodictive density
operator is then just
\begin{equation}
    \hat\rho^{\text{retr}}_{_{j,S}}(\tau)=\frac{\hat
    B(\tau)}{2x(\tau)},\label{51a}
\end{equation}
which has unit trace as required.

\subsection{Some detection events}

As a check of the retrodictive density
operator (\ref{51a}) obtained from our general retrodictive master
equation, we now calculate its matrix elements for some specific
detection events. These can be compared with corresponding
matrix elements in \cite{atom} calculated by a quite different method
involving solutions of the predictive master equation and
specific measurement POM elements.

As our first example, suppose the atom is
detected at time $\tau$ = 0 in the excited state
$|e\rangle\langle e|$. The measurement device operator corresponding
to this detection event is proportional to $|e\rangle\langle e|$,
which is also the retrodictive
density operator at this time. Thus $\hat B(0)$ $\propto$
$|e\rangle\langle e|$. We note that it is not necessary to say
whether or not the measurement device operator is an element of a POM.
By writing $|e\rangle\langle e|$ as $(1+\hat\sigma_{3})/2 $ we see
that $u(0)$ and $v(0)$
are both zero and $w(0)$ = $x(0)$. With these values we find that
\begin{equation}
   u(\tau)=0\label{52a}
\end{equation}
\begin{equation}
   v(\tau)=w(0)\exp(-3\gamma\tau/2)
   V\Omega^{-1}\sin(\Omega\tau)\label{53a}
\end{equation}
\begin{equation}
   w(\tau)=w(0)\exp(-3\gamma\tau/2)[\cos(\Omega\tau)
   -\gamma(2\Omega)^{-1}\sin(\Omega\tau)]\label{54a}
\end{equation}
and
\begin{eqnarray}
   x(\tau)&=&\frac{w(0)}{2\gamma^{2}+V^{2}}\{V^{2}
 -2\gamma\exp(-3\gamma\tau/2) \nonumber
 \\
    &&\times [\frac{(\gamma^{2}+2V^{2})}{2\Omega}
    \sin(\Omega\tau)
    -\gamma\cos(\Omega\tau)]\}.\label{55a}
\end{eqnarray}
The matrix elements of $\hat\rho^{\text{retr}}_{_{j,S}}(\tau)$ are
easily found from (\ref{51a}) and (\ref{38a}) to be
\begin{eqnarray}
   \langle e|\hat\rho^{\text{retr}}_{_{j,S}}(\tau)|g\rangle
   =\frac{u(\tau)-iv(\tau)}{2x(\tau)}
   \\
   \langle e|\hat\rho^{\text{retr}}_{_{j,S}}(\tau)|e\rangle
   =\frac{w(\tau)+x(\tau)}{2x(\tau)}\label{57a}
\end{eqnarray}
with $\langle g|\hat\rho^{\text{retr}}_{_{j,S}}(\tau)|e\rangle$ =
$\langle e|\hat\rho^{\text{retr}}_{_{j,S}}(\tau)|g\rangle^{*}$ and
$\langle g|\hat\rho^{\text{retr}}_{_{j,S}}(\tau)|g\rangle$ =
$1 - \langle e|\hat\rho^{\text{retr}}_{_{j,S}}(\tau)|e\rangle$.
From (\ref{52a}), (\ref{53a}) and (\ref{54a}) we obtain

\begin{equation}
   \langle e|\hat\rho^{\text{retr}}_{_{j,S}}(\tau)|g\rangle
   =\frac{-iVw(0)}{2x(\tau)\Omega}\exp(-3\gamma\tau/2)\sin(\Omega\tau)
   \label{58a}
\end{equation}
\begin{eqnarray}
   \langle e|\hat\rho^{\text{retr}}_{_{j,S}}(\tau)|e\rangle
   &=&\frac{w(0)}{2x(\tau)(V^{2}+2\gamma^{2})} \{V^{2}\nonumber
   \\&&\mbox{}+\exp(-3\gamma\tau/2)[(V^{2}+4\gamma^{2})
   \cos(\Omega\tau)\nonumber
\\
   &&\mbox{}-\frac{\gamma}{2\Omega}(5V^{2}+4\gamma^{2})\sin(\Omega\tau)]\}.
   \label{59a}
\end{eqnarray}
These are identical to the corresponding expressions derived in
\cite{atom} except that here we have $x(\tau)/w(0)$ in place of the
normalization factor $N$. As we already have an expression for
this in (\ref{55a}) there is no need to calculate it separately
as we had to in \cite{atom}. These results therefore not only confirm
the validity of
our new method but also display its advantages.

The Rabi oscillations are clearly evident in both the diagonal and
off-diagonal elements of the density matrix. In the limit of
long $\tau$, that is, in the distant past, the off-diagonal elements
vanish and both diagonal elements tend to 1/2. This is the density
matrix describing the no-information state discussed earlier. The
actual state the atom was prepared in is thus essentially unretrodictable for
these long times. This contrasts with the predictive density matrix,
which also exhibits Rabi oscillations for short times but for long
times in the future tends to an equilibrium state which is not the
no-information state, instead it is determined by the relative value of
$V$ and $\gamma$ \cite{book}.

If the atom is detected in its ground state, we simply take $u(0)$ and $v(0)$
as zero and $w(0)$ = $-x(0)$ and obtain a slight variation on
the above elements of the retrodictive density matrix.

If the atom is detected in the superposition state
$2^{-1/2}(|e\rangle +|g\rangle)$, the measurement device operator
is proportional to $(1+\hat\sigma_{1})$, so $v(0)$ and $w(0)$
are zero and $u(0)$ = $x(0)$. Then we obtain simply
\begin{eqnarray}
   u(\tau)&=&u(0)\exp (-\gamma\tau)\label{60a}
\\
   v(\tau)&=&0\label{61a}
\\
   w(\tau)&=&0\label{62a}
\\
   x(\tau)&=&u(0).\label{63a}
\end{eqnarray}
The retrodictive density operator is then
easily found from (\ref{51a}) and (\ref{38a}) to be
\begin{equation}
    \hat\rho^{\text{retr}}_{_{j,S}}(\tau)=
    [1+\hat\sigma_{1}\exp(-\gamma\tau)]/2.\label{64a}
\end{equation}
This has matrix elements
\begin{eqnarray}
   \langle e|\hat\rho^{\text{retr}}_{_{j,S}}(\tau)|g\rangle
   &=&[\exp(-\gamma\tau)]/2
   \\
   \langle e|\hat\rho^{\text{retr}}_{_{j,S}}(\tau)|e\rangle
   &=&1/2 \label{66a}
\end{eqnarray}
which are precisely those calculated in \cite{atom}.  Because of the
phase of the detected state there are no Rabi oscillations, the
density operator (\ref{64a}) simply decays to the no-information state
in the infinite past. To obtain Rabi oscillations in retrodiction from
a detected state that is an equal superposition of $|e\rangle$ and
$|g\rangle$, we could instead use a measuring device that
detects the state $2^{-1/2}(|e\rangle +i|g\rangle)$,
for which the measurement device operator
is $(1+\hat\sigma_{2})/2$ so $v(0)$ is not zero. Some of the
oscillatory terms in (\ref{48a}), (\ref{49a}) and (\ref{50a}) are
then retained.

\subsection{Preparation probabilities}

The elements of the retrodictive density matrix, diagonal or
off-diagonal, do not
translate into preparation probabilities until we specify the
operation of the preparation device, that is, the values of
$\hat\Lambda_{i,S}$ for use with (\ref{22a}). The simplest calculations
of preparation probabilities are those in the infinite-$\tau$ limit.
As we have seen, in this limit the retrodictive density operator
becomes proportional to the unit operator. Substituting this into
(\ref{22a}) yields
\begin{equation}
    P^{\Lambda\Gamma}(i|j)=\frac{\text{Tr}_{S}
    \hat\Lambda_{i,S}}
    {\text{Tr}_{S}\hat\Lambda_{S}},
    \label{67a}
\end{equation}
which from (\ref{A5}) and (\ref{A7}) is seen to be just the {\em a
priori} probability for the preparation event {\em i}. Thus no further
information has been gained from knowing the measurement event.

We consider a preparation device which can prepare the two-level atom
in a pure state $|i\rangle$ that is either $|e\rangle$, $|g\rangle$ or some
superposition of the two. One might envisage a device that can apply
pulses of resonant coherent radiation of various durations and phases
to an atom in its ground state. The application of a $\pi$ pulse would
prepare the atom its the excited state and a $\pi/2$ pulse with
appropriate phase would prepare the state $2^{-1/2}(|e\rangle
+|g\rangle)$. A zero pulse would prepare the atom in the ground
state. The operation of such a device is described by a set of
preparation device operators, each of which represents a possible
prepared state and incorporates the {\em a priori} probability for
that state to be prepared. That is $\hat\Lambda_{i,S}$ is
proportional to $P^{\Lambda}(i)|i\rangle\langle i|$. Clearly, even
if the prepared states are restricted to $|e\rangle$ and $|g\rangle$
the preparation device
operators will not be the elements of a POM unless these are
produced with equal {\em a priori} probabilities. If the device can
only prepare the states $2^{-1/2}(|e\rangle+|g\rangle)$ and
$|g\rangle$, the preparation device operators are not proportional to
the elements of a POM,
whatever each {\em a priori} probability is.  This is the situation for
most preparation devices and can be contrasted with measurement
devices, very many of whose operations can be described by
a POM.

To be specific, let us assume that the preparation device is
unbiased, preparing the atom in states $|e\rangle$ and $|g\rangle$
with equal probability. Then
$\hat\Lambda_{e,S}$ = $|e\rangle\langle e|/2$
and $\hat\Lambda_{g,S}$ = $|g\rangle\langle g|/2$ and the preparation
probabilities for states $|e\rangle$ and $|g\rangle$ become, from
(\ref{22a}), simply the
diagonal elements $\langle
e|\hat\rho^{\text{retr}}_{_{j,S}}(\tau)|e\rangle$ and
$\langle g|\hat\rho^{\text{retr}}_{_{j,S}}(\tau)|g\rangle$. On the
other hand, for a {\em biased} device that prepares the atom in states
$|e\rangle$ and $|g\rangle$ with {\em a priori} probabilities $p$
and $1 - p$ respectively, $\hat\Lambda_{e,S}$ and $\hat\Lambda_{g,S}$
are $p|e\rangle\langle e|$ and $(1 - p)|g\rangle\langle g|$. Then the
probability that the atom was prepared in the excited state if the
measurement is {\em j} is
\begin{equation}
    P^{\Lambda\Gamma}(e|j)=\frac{p\langle
e|\hat\rho^{\text{retr}}_{_{j,S}}(\tau)|e\rangle}
    {p\langle e|\hat\rho^{\text{retr}}_{_{j,S}}(\tau)|e\rangle+
    (1-p)\langle
    g|\hat\rho^{\text{retr}}_{_{j,S}}(\tau)|g\rangle}.\label{68a}
\end{equation}
In the limit of long $\tau$, this expression becomes equal to $p$,
the {\em a priori} probability that the atom was prepared
in the excited state, as discussed above. For short times $\tau$, the
dependence of the probability (\ref {68a}) on $\tau$ will exhibit an
oscillatory behavior depending on the state detected.

A simple but interesting case is for an unbiased preparation device that
prepares the states $|+\rangle$ = $2^{-1/2}(|e\rangle+|g\rangle)$
and $|-\rangle$ = $2^{-1/2}(|e\rangle-|g\rangle)$ with equal
probability. The preparation device operators are $|+\rangle\langle+|/2$
and $|-\rangle\langle -|/2$. The preparation probabilities for
$|+\rangle$ and $|-\rangle$ if the measurement event is {\em j} are,
from (\ref{22a}), just
$\langle +|\hat\rho^{\text{retr}}_{_{j,S}}(\tau)|+\rangle$ and
$\langle -|\hat\rho^{\text{retr}}_{_{j,S}}(\tau)|-\rangle$ respectively.
For the case discussed earlier in which the atom is detected in the
superposition state $2^{-1/2}(|e\rangle +|g\rangle)$ and so the
retrodictive density operator is given by (\ref{64a}), we find that these
probabilities are just $[1+\exp(-\gamma\tau)]/2$ and $[1-\exp(-\gamma\tau)]/2$
respectively. The values of the probabilities change from unity and zero for
small $\tau$ to both being one
half for large $\tau$ and do not exhibit oscillations at any time.
Thus the decay from the measured state to the no-information state is
shown explicitly as the only change with time involved.

The above examples illustrate the procedure involved and the preparation
probabilities for preparation devices with other
operations are easily calculated from (\ref {22a}).

\section{Conclusion}

The emerging importance of quantum communication and quantum
cryptography has made it worthwhile to re-investigate quantum
retrodiction as a means of solving the basic quantum communication
problem. This involves calculating the probabilities that
particular states were prepared when various states are detected.
Where original investigations of quantum retrodiction were in terms of
closed systems, quantum communication is more likely to involve open
systems because of interaction with the environment. For closed
systems the evolution is unitary, so the backward-time evolution
equation for the state from measurement to preparation is just the
simple inverse of the forward-time equation. For open systems the
situation is not so simple. When a system is
weakly coupled to a large environment, the
forward-time evolution is given by a master equation based on
knowledge of the prepared state of the system and the initial state
of the environment. In the retrodictive situation, the measured state
of the system is known, but not the final state of the environment
because this is not measured. Instead, the initial state of the
environment is known. Thus the backward-time evolution equation
will not be the simple inverse of the forward-time master equation. In
this paper we have derived the general, that is Lindblad, form of the
backward-time, or retrodictive, master equation. We also prove that
the retrodictive density operator remains non-negative definite at
all times between measurement and preparation. As well as confirming
the legitimacy of the master equation as an equation for the a retrodictive
density operator, the non-negativity also allows us to interpret
the evolution plus the measurement in terms of the operation of another
measurement device. This shows that we can consider the measurement,
or collapse of the state, as taking place immediately after the
preparation if we wish. Indeed it is possible to have a consistent
interpretation of the preparation-measurement process with the
collapse taking place at any time between preparation and measurement,
showing the somewhat arbitrary nature of the concept. The vanishing of
the time derivative of (\ref{25a}), or in general of (\ref{24a}),
effectively incorporates the invariance of the physically measurable
probability $P^{\Lambda\Gamma}(j|i)$ with choice of collapse time.

The retrodictive master
equation is not in general linear, underlining the fact that it is
not the simple inverse of the predictive master equation. We have
shown, however, that it can always be
linearized by means of a suitable substitution, allowing analytical
solutions for particular systems. As an explicit example, we have
considered retrodiction for an atom driven by a coherent optical
field and have solved the master equation
for the general retrodictive density operator at all times between
measurement and preparation. This allows us to select any particular
measurement event we wish and easily find the probabilities of various
possible preparation events based on this knowledge. Where the results
overlap with previous calculations based on an entirely different, and
less direct, method \cite{atom} there is perfect agreement, which
serves as confirmation of the validity of our new approach.

As we have discussed previously \cite{usBay}, preparation probabilities can
always be found from predictive evolution plus inference by means of
Bayes' theorem. Retrodictive quantum mechanics, however, offers a more
direct and often simpler means of finding these probabilities as well
as giving a different insight into the interpretation of the quantum
state. In this paper we have extended the theory of quantum
retrodiction for open systems by deriving of a general retrodictive
master equation.

\begin{acknowledgments}
We thank the United Kingdom
Engineering and Physical Sciences Research Council, the Australian
Research Council, the Royal Society of Edinburgh and the Scottish
Executive Education and Lifelong Learning Department for financial support.
We also thank N. L\"utkenhaus for useful discussions.
\end{acknowledgments}


\begin{thebibliography}{99}
\bibitem{von} J. von Neumann, {\em Mathematical Foundations of Quantum
Mechanics} (Princeton University Press, Princeton, 1955).
\bibitem{hel} C. W. Helstrom, {\em Quantum Detection and Estimation Theory}
(Academic Press, New York, 1976).
\bibitem{Ahar} Y. Aharonov, P. G. Bergman and J. L. Lebowitz, Phys. Rev.
{\bf 134}, B1410 (1964).
\bibitem{usBay} S. M. Barnett, D. T. Pegg, and J. Jeffers, J. Mod. Opt.
{\bf 47}, 1779 (2000).
\bibitem{lots} R. H. Penfield, Am. J. Phys. {\bf 34}, 422 (1966); Y. Aharonov
and D. Z. Albert, Phys. Rev. D {\bf 29}, 223 (1984);
Y. Aharonov and D. Z. Albert, Phys. Rev. D {\bf 29}, 228 (1984);
Y. Aharonov and L. Vaidman, J. Phys. A: Math. Gen. {\bf 24}, 2315 (1991);
D. T. Pegg and S. M. Barnett, Quantum and Semiclass. Opt. {\bf 1}, 442
(1999); F. J. Belinfante, {\em  Measurements and Time Reversal in Objective
Quantum Theory} (Pergamon Press, Oxford,1975).
S. M. Barnett, D. T. Pegg, J. Jeffers, O. Jedrkiewicz and R. Loudon  Phys.
Rev. A  {\bf 62}, 022313 (2000).
\bibitem{2} Quantum Communication, Computing and Measurement, edited by O.
Hirota, A. S. Holevo and C. M. Caves
(Plenum, New York, 1997).
\bibitem{3} S. J. D. Phoenix and P. D. Townsend, Contemp. Phys. {\bf 36},
165 (1995) and references therein; D. Bouwmeester, A. Ekert and A. Zielinger
(eds.), {\it The Physics of Quantum Information} (Springer, Berlin, 2000) and
references therein.
\bibitem{PRL} S. M. Barnett, D. T. Pegg, J. Jeffers, and
O. Jedrkiewicz, Phys. Rev. Lett. {\bf 86}, 2455 (2001).
\bibitem{prep} D. T. Pegg, S. M. Barnett, and J. Jeffers, J. Mod.
Opt. (in press).
\bibitem{Boas} M. L. Boas, {\em Mathematical Methods in the Physical
Sciences} (Academic Press, New York, 1983), p. 695.
\bibitem{Bayes} G. E. P. Box and G. C. Tiao, {\em Bayesian Inference in
Statistical Analysis}
( Addison-Wesley, Sydney, 1973), p. 10.
\bibitem{BB84} C. H. Bennett and G. Brassard, {\em Proceedings of the
IEEE International Conference on Computers, Systems and Signal
Processing} (Bangalore, 1984), p. 175.
\bibitem{book} See, for example, S. M. Barnett and P. M. Radmore {\em
Methods in Theoretical Quantum Optics}
(Oxford University Press, Oxford, 1997).
\bibitem{11} G. Lindblad, Commun. Math. Phys. {\bf 48}, 119 (1976); S.
Stenholm, in {\em Quantum Dynamics of Simple Systems}, edited by G.-L.
Oppo, S. M. Barnett, E. Riis, and M. Wilkinson (SUSSP Publications,
Edinburgh, 1996) p. 299.
\bibitem{Menzel} See, for example, D. H. Menzel, {\em Fundamental
Formulas of Physics} (Dover Publications, New York, 1960) p. 36.
\bibitem{atom} S. M. Barnett, D. T. Pegg, J. Jeffers, and O. Jedrkiewicz,
J. Phys. B, At. Mol. Opt. Phys. {\bf 33}, 3047 (2000).

\end{thebibliography}

\end{document}